\documentclass[aps,reprint,amsmath,amssymb,]{revtex4-2}

\usepackage{graphicx}
\usepackage{dcolumn}
\usepackage{bm}

\usepackage[utf8]{inputenc}
\usepackage[T1]{fontenc}
\usepackage{mathptmx}

\def\sub#1{_{\rm #1}}  
\def\U#1{\,{\rm #1}}

\begin{document}


\title[]{Storage and retrieval of electromagnetic waves in a metasurface based on bound states in the continuum by conductivity modulation}


\author{Toshihiro Nakanishi}
 \affiliation{Department of Electronic Science and Engineering, Kyoto University, Kyoto 615-8510, Japan}




\begin{abstract}
 In this study, we develop a time-varying metasurface 
 based on the bound states in the continuum (BIC) with variable conductors,
 to store electromagnetic waves.
 The storage and retrieval of electromagnetic waves are demonstrated numerically
 through dynamic switching between quasi-BIC and BIC states
 by modulating the variable conductors.
 The storage efficiency exhibits oscillatory behavior with respect to the timing of
 storage and retrieval.
 These behaviors can be attributed to the interference of
 a resonant mode and a static mode that is formed by direct current.
In addition, the storage efficiency of a single-layer metasurface can reach
35\% under ideal conditions. 
\end{abstract} 


\maketitle


Metamaterials are artificial media composed of sub-wavelength structures,
whereas metasurfaces are two-dimensional metamaterials.
They have been developed to extend the tunability of  medium parameters.
The permeability of metamaterials covers a wide range,
including negative values applied to negative refraction \cite{Pendry.1999,Shelby.2001}.
Moreover, the spatial modulation of permittivity and permeability has led to 
the development of transformation optics \cite{leonhardt2012geometry}, 
which realizes an invisible cloak \cite{Schurig.2006}.
Recently, the development of time-varying media and metamaterials,
which have properties that can be controlled over time,
has attracted considerable attention for achieving novel electromagnetic phenomena
\cite{Shlivinski.2018,Galiffi.2020,Wang.2020,Marini.2021,Tirole.2023}
such as non-reciprocal amplification \cite{Galiffi.2019},
Fresnel drag \cite{Huidobro.2019},
anti-reflection coatings in the time domain \cite{PachecoPena.2020, Ramaccia.2020},
frequency conversion through time refraction
\cite{Akbarzadeh.2018,Lee.2018,Shcherbakov.2019,Karl.2020,Zhou.2020,Miyamaru.2021},
temporal reflection \cite{Moussa.2023}.
The storage and retrieval of electromagnetic waves were realized 
by the temporal modulation of the medium parameters.
This was experimentally demonstrated in atomic systems
that realize the nonlinear optical effect referred to as the electromagnetically-induced
transparency (EIT) effect \cite{Phillips.2001,Liu.2001}.
The EIT effect was extended to the field of metamaterials
\cite{Fedotov.2007,Zhang.2008,Papasimakis.2008,Nakanishi.2015}.
The storage and retrieval of electromagnetic waves were demonstrated within the microwave region
using metamaterials integrated with nonlinear capacitors \cite{Nakanishi.2013,Nakanishi.2018}.

An  alternative approach is available for realizing 
the storage and retrieval of electromagnetic waves
by the dynamic modulation of bound states in the continuum (BIC).
The BIC, which is localized modes in a continuous spectrum,
was first introduced in the field of quantum mechanics
and then extended to general classical waves, including electromagnetic waves \cite{Hsu.2016}.
The resonant mode of the ideal BIC, which is completely decoupled from the propagating waves,
is typically implemented using symmetric structures.
The breaking of the structural symmetry leads to interactions with the propagating waves
\cite{Campione.2016,Tittl.2018,Tuz.2018}.
The resonances in the BIC and quasi-BIC have
extremely high-quality factors due to small radiation losses
\cite{Rybin.2017,Kodigala.2017,Carletti.2018}.
The storage of electromagnetic waves can be achieved
by switching between a quasi-BIC state that is weakly coupled to the propagating waves and
a BIC state that is a completely localized. 
The light storage in photonic crystal waveguide was theoretically demonstrated by utilizing Kerr effect \cite{Bulgakov_Pichugin_Sadreev_2015}. 
A recent study theoretically demonstrated the BIC-based storage of spherical light waves
in a single core-shell resonator
by modulating the outer-shell permittivity to approximately zero \cite{Hayran.2021}.

In this study, we develop a metasurface
composed of periodic planar structures, which are composed of metals and variable conductors
with states that can be switched between quasi-BIC and BIC.
The storage of plane waves in the BIC state is numerically demonstrated
by modulating the conductivity of the variable conductors.
Furthermore, the storage efficiency based on a numerical analysis with respect to various storage and retrieval timings, which shows interference phenomena,
reveals that energy is stored in an oscillating mode and a
static mode that is formed by direct current (DC), 
which is unique to a time-varying system implemented using conductivity modulation
\cite{Qu.2018,Miyamaru.2021,Nakata.2022}.

\begin{figure}[b] 
  \centering\includegraphics[]{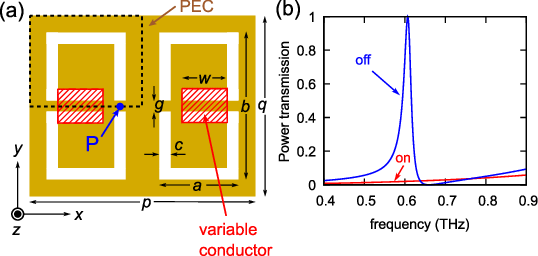}  
  \caption{(a) Unit cell of the metasurface. Point P at the center of the cSRR gaps is located 1\,\textmu m above the metal surface. (b) Transmission spectra for the off-state
  and on-state.}
  \label{unit-trans}
  \end{figure}

Figure~\ref{unit-trans}(a) presents the unit structure of the metasurface,
which is composed of a pair of complimentary split ring resonators (cSRRs)
with a metallic sheet.
The metasurface includes variable conductors as active elements
with electric properties that can be dynamically controlled between the insulating 
and high-conductivity states.
Asymmetric cSRRs are realized for variable conductors in the insulating state (off-state),
and symmetric cSRRs are realized for those in the high-conducting state (on-state), which virtually 
fill the insulating holes with length $w$.
The  parameters are designed as $p=280$\,\textmu m, $q=140$\,\textmu m,
$a=80$\,\textmu m, $b=120$\,\textmu m, $c=8$\,\textmu m, 
$g=4$\,\textmu m, and $w=39$\,\textmu m
for operation in the terahertz regime. 
We assume that the metallic sheet is a perfect electric conductor (PEC) with zero thickness.
In addition the variable conductors have a thickness of 1\,\textmu m and a conductivity of 
$\sigma=0$ for the off-state and $\sigma=10^{10}\U{S/m}$ for the on-state,
thus practically acting as a lossless conductor or PEC.
The transmission spectra are calculated using  electromagnetic simulation software, CST Studio Suite. 
To reduce the simulation cost, we focus on the dotted region obtained by dividing the unit structure 
into four equal parts, as shown in Fig.~\ref{unit-trans}(a).
The PEC boundaries are imposed on the two sides of the dotted region normal to the $x$ axis
and perfect magnetic conductor boundaries are imposed normal to the $y$ axis.
Under these conditions, the normal incidence of $x$-polarized waves can be simulated for periodically arranged unit structures owing to the symmetry of the structure.  
The metasurface is placed in the middle of a simulation space with a length $L=1000$\,\textmu m.
See Sec.~1 in Supplement 1 for further details.

Figure~\ref{unit-trans}(b) represents the transmission spectra for the off-state
and on-state, which were calculated using a transient solver.
The off-state (insulating state) exhibits a sharp transmission peak at $0.607\U{THz}$,
which results from the excitation of the resonance in the cSRR structure.
In contrast, for the on-state, the transmission is low over the entire spectral region.
In this case, the resonant state in the metasurface is effectively decoupled from 
the propagating waves, and a BIC is realized.
Transition between the BIC state in the symmetric structure 
and the quasi-BIC state in the symmetry-broken structure is a typical characteristic of BIC-based metamaterials \cite{Koshelev.2018}.
We identify the resonant mode for the on-state, which cannot be directly excited by
the incident waves, using an eigenmode solver, and
the resonant frequencies are estimated to be $0.757\U{THz}$ for the BIC state (on-state)
and $0.604\U{THz}$ for the quasi-BIC state (off-state).
This is in good agreement with the transmission peak shown in Fig.~\ref{unit-trans}(b).

The storage of electromagnetic waves in the proposed metasurface can be 
demonstrated in the following procedure.
First, the metamaterial is prepared in a quasi-BIC state to receive  propagating electromagnetic waves, which are tuned to excite the resonance of the cSRRs. 
Second, the state is switched into the BIC state, and the energy is captured in the metasurface
owing to the decoupling of the resonant mode and the outer space.
Finally, the state returns to the quasi-BIC state again,
and the energy captured in the metasurface is released as  propagating waves through 
the coupling.

\begin{figure}[t]
\centering\includegraphics[]{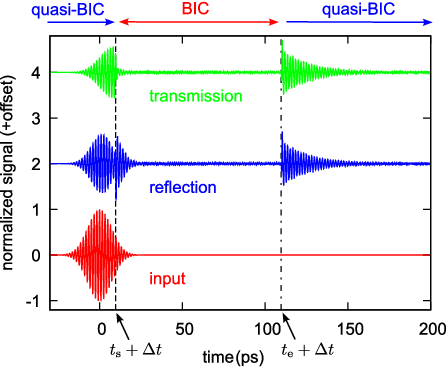} 
\caption{Input (bottom), reflection (middle), and transmission 
(top) electric fields for $t\sub{w}=10\U{ps}$ , $t\sub{s}=8\U{ps}$, and $t\sub{e}=108\U{ps}$.}
\label{waveform}
\end{figure}

The input electric field at $z=-L/2$ 
is a Gaussian pulse with a temporal duration of $t\sub{w}=10\U{ps}$ 
and carrier frequency of $0.607\U{THz}$, 
which is tuned to the transmission peak of the quasi-BIC state, or off-state,
as shown at the bottom of Fig.~\ref{waveform}.
The time origin $t=0$ is set at the center of the Gaussian pulse.
The metasurface changes from the quasi-BIC state 
to the BIC state at $t\sub{s}=8\U{ps}$ with a transition time of $10\U{fs}$,
and returns to its original state (quasi-BIC state) at $t\sub{e}=108\U{ps}$
with a transition time of $10\U{fs}$.
In this case, the storage duration is expected to be $\tau=t\sub{e}-t\sub{s}=100\U{ps}$.
The calculated reflection signal at the input port $z=-L/2$ and
transmission signal at $z=+L/2$ are indicated at the middle and 
top of Fig.~\ref{waveform}, respectively.
The dashed and dash-dotted lines represent $t=t\sub{s}+\Delta t$ and $t=t\sub{e}+\Delta t$,
where $\Delta t=L/2 c_0=1.67\U{ps}$ denotes the propagation time between the metasurface and each port
for the speed of light in a vacuum $c_0$.
It is reasonable that both the transmitted and 
the reflected signals exhibit abrupt changes
at $t=t\sub{s}+\Delta t$ and $t=t\sub{e}+\Delta t$
due to the propagation time $\Delta t$ from the metasurface $z=0$ to each port $z=\pm L/2$.
Moreover,
each signal is suppressed, except for the transient reflection signal immediately after $t\sub{s}+\Delta t$.
This is because the energy is captured in the BIC state and decoupled from the outer space.
Immediately after $t\sub{e}+\Delta t$, the transmission and reflection ports receive 
completely identical signals,
which implies that the captured energy is equally released 
from the quasi-BIC state in both directions 
through the coupling between the resonant mode and propagation modes.
The electric field distributions in each phase are provided in Sec.~2 in Supplement 1.

\begin{figure}[t]
	\centering\includegraphics[]{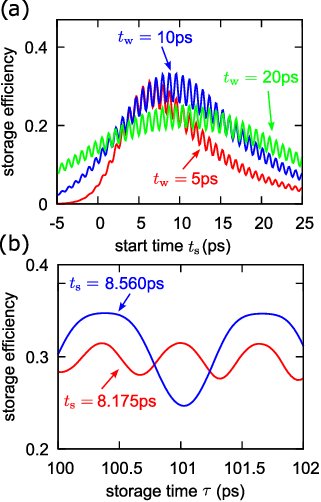} 
	\caption{(a) Storage efficiency as a function of $t\sub{s}$ for
	$t\sub{w}=5\U{ps}, 10\U{ps}$, and $20\U{ps}$.
	(b) Storage efficiency as a function of $\tau$ for
	$t\sub{s}=8.175\U{ps}$ and $8.560\U{ps}$.}
	\label{efficiency} 
\end{figure}

The total energy released from the metasurface can be estimated by integrating
the squares of the reflection and transmission signals after $t_{\rm e}$.
The storage efficiency $\eta$ can be obtained by dividing it by the input energy,
which can be obtained by integrating the square of the input signal.
The storage efficiency depends on the input signal parameters
and modulation timings.
We computed the dependence of the storage efficiency on the start time $t\sub{s}$
for the following pulse widths: $t\sub{w}=5, 10$ and $20\U{ps}$.
The results are shown in Fig.~\ref{efficiency}(a) for a storage time of $\tau=100\U{ps}$.
It is deduced that the case of $t\sub{w}=10\U{ps}$ is the optimal among these three.
The lifetime of the quasi-BIC state is estimated to be $7.9\U{ps}$,
which is obtained from transmission or reflection signals after $t=t\sub{e}$,
and it is close to the optimal pulse width $t\sub{w}=10\U{ps}$. 
This is because shorter and longer pulses 
cannot effectively excite the resonance in the quasi-BIC state.
In all cases, the storage efficiency oscillates with respect to the start time $t\sub{s}$
when the metasurface changes from the quasi-BIC state to the BIC state.
For example, the efficiency reaches an optimal value at $t\sub{s}=8.560\U{ps}$
and a local minimum at $t\sub{s}=8.175\U{ps}$ for $t\sub{w}=10\U{ps}$.
The oscillation period is estimated as $\Delta t\sub{s}=2(8.560\U{ps}-8.175\U{ps})=0.77\U{ps}$, 
which is close to half the resonance period of the cSRRs in the quasi-BIC state 
$1/0.607\U{THz}=1.65\U{ps}$.
This  suggests that the storage efficiency depends on the phase of the resonance relative to $t\sub{s}$, as detailed further in the manuscript.
Thereafter, we fixed the pulse width at $t\sub{w}=10\U{ps}$ and
calculated the storage efficiency as a function of storage time $\tau$ for 
$t\sub{s}=8.175\U{ps}$ and $t\sub{s}=8.560\U{ps}$.
(The output power including uncaptured component is provided in Sec.~3 in Supplement 1.)
Both cases exhibit oscillatory behaviors as shown in Fig.~\ref{efficiency}(b).
However, they exhibit different characteristic periods,
which are $0.65\U{ps}$ for $t\sub{s}=8.175\U{ps}$ and $1.3\U{ps}$ for $t\sub{s}=8.560\U{ps}$.
The optimal storage efficiency is 0.35 at $\tau=100.375\U{ps}$ for $t\sub{s}=8.560\U{ps}$.

\begin{figure}[t]
	\centering\includegraphics[]{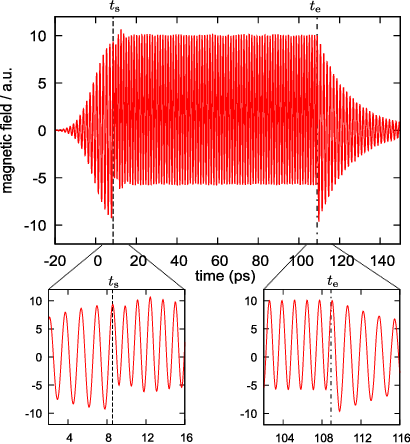}  
	\caption{Magnetic field in $y$-direction at Point P for optimized parameters,
	$t\sub{s}=8.560\U{ps}$ and $t\sub{e}=108.935\U{ps}$ ($\tau=100.375\U{ps}$ ).
	The dashed and dash-dotted lines  indicate $t=t\sub{s}$ and $t=t\sub{e}$, respectively.}
	\label{h-field}
	\end{figure}

To explain the oscillatory behaviors shown in Figs.~\ref{efficiency}(a) and (b),
the magnetic field at Point P defined in Fig.~\ref{unit-trans}(a),
which is located 1\,\textmu m above the metal surface, was calculated 
as for the optimal case, i.e., the start time $t\sub{s}=8.560\U{ps}$ 
and a storage time $\tau=100.375\U{ps}$.
The $y$ component of the magnetic field is shown at the top of Fig.~\ref{h-field},
where $t=t\sub{s}(=8.560\U{ps})$ and $t=t\sub{e}(=108.935\U{ps})$ are denoted by the dashed and dash-dotted lines, respectively.
Magnified views around $t=t\sub{s}$ and $t=t\sub{e}$ are shown at the bottom left and bottom right of 
 Fig.~\ref{h-field}.
(The calculation in the presence of ohmic loss is provided in Sec.~4 in Supplement 1.)
Steady oscillation is maintained during 
$t\sub{s} < t < t\sub{e}$, thus indicating that 
the radiation loss of the BIC state is negligible. 
It is found that the magnitude of the magnetic field is maximized at the start time 
$t=t\sub{s}$ and the end time $t=t\sub{e}$ of the storage process.
Hence, storage efficiency is maximized by switching the states of the variable conductors
when the magnitude of the magnetic field is maximized.
This can be explained as follows.
The cSRRs can be regarded as inductor-capacitor circuits,
where the inductor is formed by the metal and the capacitor by the holes.
During the resonance process,
when the energy stored in the inductors is maximized, that stored in the capacitors
is minimized.
This is the most appropriate moment to switch the variable conductors from insulators to metals
because the capacitors do not accumulate charges that can vanish
by short-circuiting the holes forming the capacitances.
This is why the start time is optimized at $t\sub{s}=8.560\U{ps}$, 
when the magnitude of the magnetic field is maximized.
The magnitude of the magnetic field reaches its maximum at half the interval of the resonance period.
This is consistent with the period of the efficiency oscillation, 
which is close to half of the resonance period of the cSRRs in the quasi-BIC state, 
as illustrated in Fig.~\ref{efficiency}(a).
During the storage period $t\sub{s} < t < t\sub{e}$,
the magnetic field oscillates at $0.78\U{THz}$,
which is up-shifted from $0.607\U{THz}$ through the switching and 
agrees well with the resonant frequency $0.757\U{THz}$ derived from the eigenmode analysis. 
Notably, the magnetic-field oscillation has an offset.
This is caused by the direct current (DC) in the metal and the metallized variable conductors.
The excitation of the DC mode in the BIC state can be clearly observed in Visualization 1,
which presents an animation of the magnetic field at the metasurface.
This ``DC mode'' is universally excited at the temporal boundary in a time-varying system
to satisfy the continuous conditions for the electric and magnetic fields \cite{Miyamaru.2021}.
The instant at which the field distribution at the start time $t\sub{s}$ is reproduced
is optimal for the end time $t\sub{e}$, given that no current flows in the variable conductors, and
the reverse process is induced by the transition from the BIC state to the quasi-BIC state.
The magnetic field at $t=t\sub{s}$ is reproduced at the resonance period interval
due to the presence of the DC mode.
Thus, in the case of $t\sub{s}=8.560\U{ps}$ in Fig.~\ref{efficiency}(b),
the period of the efficiency oscillation is close to 
the resonance period of the BIC state $1/(0.757\U{THz})=1.3\U{ps}$.

The case changes significantly when $t\sub{s}=8.175\U{ps}$,
where the storage efficiency assumes a local minimum as shown in Fig.~\ref{efficiency}(a).
This implies that the electric field is maximized at $t\sub{s}=8.175\U{ps}$ and
a fraction of the electric energy formed by the accumulated charges 
at the holes in the cSRR structure is dissipated 
through short circuiting. 
In contrast to the case of $t\sub{s}=8.560\U{ps}$, 
DC mode is negligible
because the magnetic field around the metasurface is minimized at $t=8.175\U{ps}$.
Without the DC mode, the maximization of the magnetic-field magnitude
when no charge accumulates in the holes 
occurs at the intervals of 
half the resonance period, i.e.,  $1/(2\times 0.757\U{THz})=0.66\U{ps}$,
which is in good agreement with an oscillation period of $0.65\U{ps}$ 
as shown in Fig.~\ref{efficiency}(b).

In summary, we  numerically demonstrated
the storage and retrieval of electromagnetic waves 
by switching between the quasi-BIC and BIC states
through the modulation of the variable conductors integrated into the metasurface. 
As demonstrated,
the energy is stored not only in the resonant mode but also in the DC mode excited 
through conductivity modulation,
which leads to a peculiar interference in the storage efficiency.
The storage efficiency of the single-layer metasurface was 35\% for the optimal case.
By increasing the number of layers,
the efficiency could be improved,
and the propagation direction could be preserved during the retrieval process
owing to constructive interference of waves emitted from each layer
\cite{Nakanishi.2013,Nakanishi.2018}.
We designed the metasurface for the operation in the terahertz regime.
However, the modulation of conductivity 
has been realized over a wide range of spectral regions using various methods,
such as diodes, photocarrier excitations in semiconductors, 
and phase change materials \cite{Yang.2022}.
Experimental demonstrations could be achieved using these active materials.
In this study, a lossless metal and variable conductors were assumed
to consider ideal conditions under which the radiation loss of the BIC state is negligible,
thus resulting in a clear observation of the interference between the resonant and DC modes.

Compared with previous methods using EIT metamaterials 
with the two resonant modes
operated in the microwave regions \cite{Nakanishi.2013, Nakanishi.2018}, 
the proposed method based on the modulation of a single resonant mode 
is significantly simpler, and an extension to a higher frequency is expected.
It was revealed  that the DC mode
plays an essential role in the operation of dynamic metamaterials with conductivity modulation.
The findings provide  new insights for future studies on time-varying optical systems.

 
This work was supported by JSPS KAKENHI [grant Numbers 20K05360 and	23K04612].
The author would like to acknowledge the helpful discussions with Prof.~Yosuke Nakata, Osaka University, 
and Prof.~Fumiaki Miyamaru, Shinshu University.






%

\clearpage

\onecolumngrid
\begin{center}
\textbf{\large Storage and retrieval of
electromagnetic waves in a
metasurface based on bound states in
the continuum by conductivity
modulation: supplemental document}
\end{center}

\setcounter{equation}{0}
\setcounter{figure}{0}
\setcounter{table}{0}
\setcounter{page}{1}
\makeatletter

\renewcommand{\thesection}{\arabic{section}}




\section{Simulation space}

\begin{figure}[htbp]
  \centering
  \includegraphics[scale=0.8]{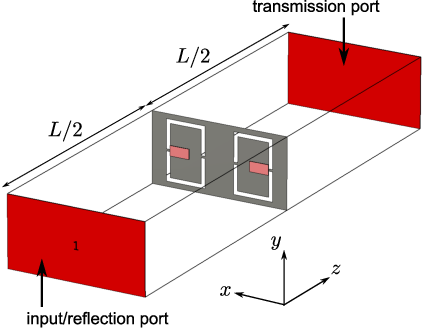}
  \caption{Simulation space and port setup.}
  \label{sfig1}
  \end{figure}

The simulation space and port setup are shown in Fig.~\ref{sfig1}.
The metasurface is placed in the middle of the simulation space with a length $L=1000$\,\textmu m.
Electromagnetic waves with $x$ polarization are sent from a port at $z=-L/2$,
which is also used to receive reflected waves, and
the other port at $z=L/2$ receives transmitted waves.

\section{Electric field distribution}

\begin{figure}[htbp]
  \centering
  \includegraphics[width=1\linewidth]{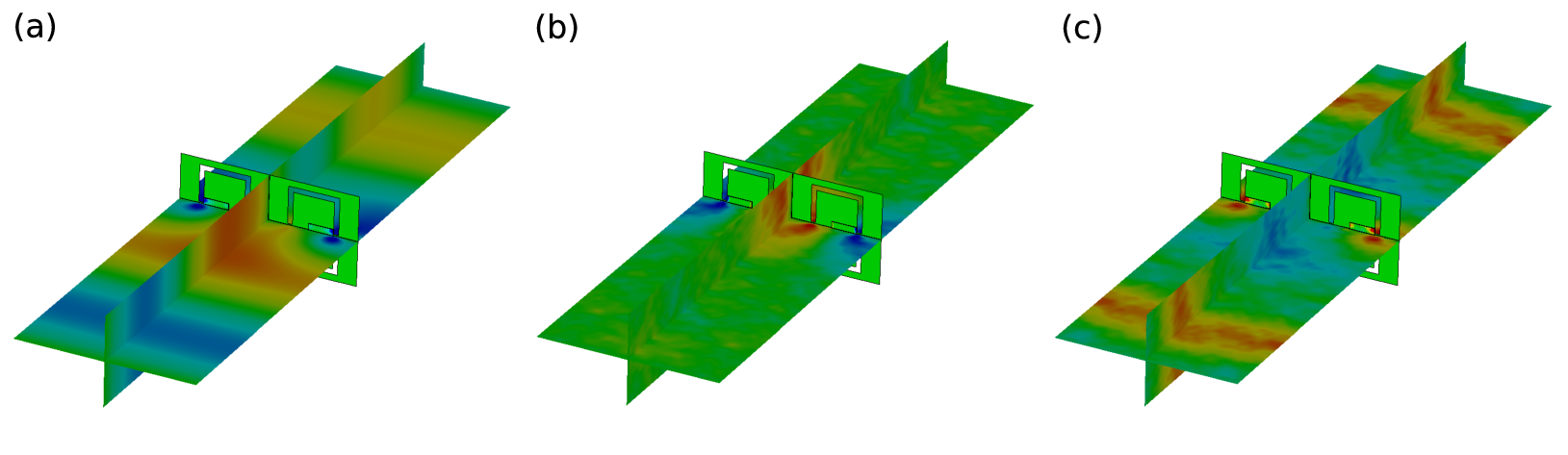}
  \caption{Electric fields in $x$ direction at 
  (a) $t=6\U{ps}$, (b) $t=50\U{ps}$, and (c) $t=110\U{ps}$
  for $t\sub{s}=8\U{ps}$ and $t\sub{e}=108\U{ps}$.}
  \label{sfig2}
  \end{figure}

Electric fields in $x$ direction for $t\sub{s}=8\U{ps}$, $t\sub{e}=108\U{ps}$ 
and $t\sub{w}=10\U{ps}$,
are shown in Fig.~\ref{sfig2}.
Propagating waves including input, reflected, and transmitted waves,
can be observed at $t=6\U{ps} \,(< t\sub{s})$ as shown in Fig.~\ref{sfig2}(a).
The electric field at $t=50\U{ps}$ in storage period in Fig.~\ref{sfig2}(b)
is localized around the metasurface,
which indicates that the energy is captured in the BIC state.
At $t=110\U{ps} \,( > t\sub{e})$ shown in Fig.~\ref{sfig2}(c), 
propagating waves emitted from the metasurface in both directions can be observed.

\section{Total output power}

\begin{figure}[htbp]
  \centering
  \includegraphics[]{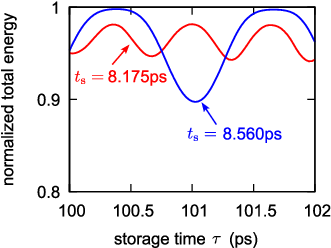}
  \caption{Total output power normalized by the input power as a function of $\tau$ for
  $t\sub{s}=8.175\U{ps}$ and $8.560\U{ps}$.}
  \label{sfig3}
  \end{figure}

Total output power is obtained by integrating the squares of transmission and reflection signals.
It includes not only stored energy observed after the modulation  $t>t\sub{e}$
but also energy before $t<t\sub{e}$, which is not captured in the BIC state. 
Figure \ref{sfig3} shows the total output power normalized 
by the input power as a function of storage time $\tau$
under the same condition as Fig.~3(b) in the main text.
The normalized output energy is always less than unity,
which indicates that the system is passive and energy is not amplified by conductivity modulation.
For $t\sub{s}=8.560\U{ps}$, the normalized output energy approaches unity at $\tau=100.375\U{ps}$,
where the storage efficiency is maximized as shown in Fig.~3(b) in the main text.
Under this condition, where the energy is almost conserved,
the electric and magnetic fields are continuously connected at the modulation timings $t\sub{s}$ and $t\sub{e}$. 
The energy conservation cannot be observed for  $t\sub{s}=8.175\U{ps}$, 
which means that the DC mode observed for $t\sub{s}=8.560\U{ps}$ contributes to the energy conservation and high storage efficiency.

\section{Resonance decay caused by ohmic loss}

Ohmic loss, which is not considered in the main text,
decreases  quality factors of the quasi-BIC and BIC states.
In order to estimate the effect of ohmic loss, 
the magnetic field in $y$-direction at Point P defined in Fig.~1 (a) in the main text
was calculated for the start time $t\sub{s}=8.560\U{ps}$,
assuming the metallic sheet with sheet resistance of $0.3\, \Omega$
and the variable conductor with conductivity of $5\times 10^5\U{S/m}$ in conducting state.
Field decay in the BIC state after the start time $t\sub{s}$ can 
be observed for both of the oscillation and DC modes.
The storage time would be limited within a few tens of picoseconds
under this condition.

    \begin{figure}[htbp]
      \centering
      \includegraphics[]{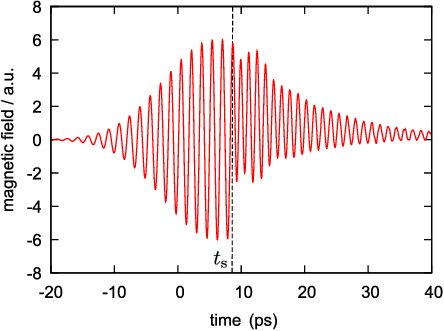}
      \caption{Magnetic field in $y$-direction at Point P for 
      $t\sub{s}=8.560\U{ps}$.
      The dashed line indicates $t=t\sub{s}$.}
      \label{sfig4}
      \end{figure}

\end{document}